\documentstyle[11pt,newpasp,twoside, epsf]{article}
\def\edcomment#1{\iffalse\marginpar{\raggedright\sl#1\/}\else\relax\fi}
\reversemarginpar

\begin{document}
\title{Alternative Solutions to Big Bang Nucleosynthesis}
\author{Hannu Kurki-Suonio}
\affil{Helsinki Institute of Physics, P.O. Box 9, FIN-00014 University of
Helsinki, Finland}

\begin{abstract}
Standard big bang nucleosynthesis (SBBN) has been
remarkably successful, and it may well be the correct and sufficient
account of what happened.  However, interest in variations
from the standard picture come from two sources:
First, big bang nucleosynthesis can be used to constrain physics of the
early universe.  Second, there may be some discrepancy between
predictions of SBBN and observations of abundances.
Various alternatives to SBBN include inhomogeneous nucleosynthesis,
nucleosynthesis with antimatter, and nonstandard neutrino physics.
\end{abstract}

\section{Introduction}

The success of standard big bang nucleosynthesis (SBBN) in predicting
the observed abundances of the light elements has led to the
widespread view that SBBN must be correct.  According to this view,
any remaining disagreements must be due to systematic errors in
observations or incorrect, or too crude, chemical evolution models.  While
this view may well be the right one,
we should not be blind to other possibilities.

I will stay within the context of the Hot Big Bang (for an alternative,
see Burbidge \& Hoyle 1998), and discuss some models of nonstandard
big bang nucleosynthesis (NSBBN).  NSBBN scenarios range
from small modifications to SBBN to a complete change in the
decisive physical phenomena, like in the late-decaying massive
particle scheme of Dimopoulos et al. (1988).

  Motivations for studying NSBBN go in two
directions.  First, the remarkable success of SBBN allows one to severely
constrain the physics of the early universe.  If one tries to change
the conditions from the standard assumptions the resulting abundances
of the light elements
differ from the observed ones.  For many modifications, BBN provides the
strongest constraints.  BBN gives also the strongest constraint on 
the single parameter of SBBN, the baryon density, usually given as the
baryon-to-photon ratio,
\begin{equation}
 \eta \equiv \frac{n_b}{n_\gamma}, \qquad \eta_{10}\equiv10^{10}\eta.
\end{equation}
Second, one may try to improve on SBBN.  From time
to time it has seemed that there might be some discrepancy
between observations and SBBN, which could then be explained by NSBBN.
In particular, there has been tension between D/H and $Y_p$
(see, e.g., Hata et al. 1995).  To relieve this tension, 
either a lower D or a lower 
${}^4$He yield has been looked for.
Also one may want to relax the SBBN bounds to $\eta$.
Other astronomical considerations have given motivation
for trying to raise the upper limit to $\eta$.
If one believes that the energy density of the universe is dominated
by vacuum energy (the cosmological constant) and accepts the newer
observations on D/H and $Y_p$ favoring somewhat larger $\eta$ within
SBBN,
this motivation largely disappears.

There is a very large body of work on NSBBN.
Extensive reviews are given by Malaney \& Mathews
(1993) and Sarkar (1996), which contain, respectively, over 500 and over
700 references.  Here I will be able to mention only a random few.

Most of the work on NSBBN can be divided into four broad classes:
\begin{enumerate}
\item Inhomogeneous BBN.  Usually this means inhomogeneity in the
baryon-to-photon ratio, $\eta$, but there are also other
possibilities, like 
inhomogeneity in the neutrino chemical potentials.
\item Nonstandard neutrino physics, e.g., additional (``sterile'')
neutrino flavors, neutrino degeneracy (asymmetry),
massive $\nu_{\tau}$, or neutrino oscillations.
\item Late-decaying ($\tau$ = 1--$10^8$ s) massive particles,
black holes, cosmic strings,  etc.
\item Time-varying fundamental constants.
\end{enumerate}
In the interest of time and space, I will discuss the
first two classes only. 

\section{Inhomogeneous Big Bang Nucleosynthesis}

The single parameter of SBBN is the baryon-to-photon ratio $\eta$,
or the density of baryonic matter.  In inhomogeneous big-bang 
nucleosynthesis (IBBN) one assumes that $\eta$ is inhomogeneous.
To get a significant effect on BBN this inhomogeneity has to be large,
$\delta\eta/\eta \ga 1$.  Since the baryons make an insignificant
contribution to the energy density at nucleosynthesis time, the total
energy density may still be essentially homogeneous.
The inhomogeneity could be caused by, e.g., first-order
phase transitions.  The distance scale of this inhomogeneity is
of crucial importance for IBBN.  Without inflation, causal physics
can only produce significant inhomogeneity at subhorizon scales
(see Table 1).

\begin{table}
\centering
\caption{The approximate temperature and horizon scale (in comoving
units) for various events in the early universe.}
\vspace{2mm}
\begin{tabular}{lcc}
\tableline
event & T & horizon  \\
\tableline
EW phase transition & 100 GeV &  $10^{-3}$ pc \\
QCD phase transition & 150 MeV & 1 pc \\
${}^4$He synthesis & 70 keV & 1 kpc \\  
\tableline
\tableline
\end{tabular}
\end{table}

Mechanisms connected with inflation can produce inhomogeneity
at any scale.  The isotropy of the cosmic microwave background (CMB)
rules out significant inhomogeneity at $\ga 10$ Mpc scales,
and it is difficult to construct an acceptable IBBN
scenario which would explain inhomogeneity in observations. 
In the usual IBBN models one considers a significantly smaller
distance scale, so that while $\eta$ is inhomogeneous during
BBN, resulting in inhomogeneous abundances at first, everything
gets mixed and becomes chemically homogeneous before or during
galaxy formation.  Thus the observable primordial abundances
are homogeneous, while different from the SBBN predictions.

The simplest version of IBBN is one where SBBN occurs with different
$\eta$ in different parts of the universe, and the yields get
mixed afterwards, so that one obtains the IBBN results by
averaging SBBN results over the $\eta$ distribution, whose
average we denote by $\bar{\eta}$.  This kind
of IBBN has a long history. Typically $Y_p$ goes up, 
${}^7$Li goes up (down for small $\bar{\eta}$), and
D goes up for large $\bar{\eta}$, and down for small $\bar{\eta}$,
compared to SBBN with $\eta = \bar{\eta}$.  Leonard \& Scherrer (1996) 
concluded that this way one can reduce the lower bound to $\eta$
from observations (in fact remove it, if arbitrary $\eta$ 
distributions are allowed), but the upper bound is essentially
unchanged from SBBN, as ${}^7$Li and ${}^4$He are overproduced for larger
$\bar{\eta}$.  The tension between D an ${}^4$He is worsened at the
large end of the SBBN acceptable range.  Thus this kind of 
modification to BBN appears undesirable.

\subsection{Small Scale Inhomogeneity and Neutron Diffusion}

The above applies to inhomogeneity
with distance scales significantly larger than the neutron diffusion
scale ($\sim 0.1$ pc).
If there is inhomogeneity at smaller scales, neutrons will
diffuse out of the high density regions resulting in an inhomogeneous
$n/p$ ratio.  Especially if this results in 
$n/p > 1$ in some regions, the consequencies for BBN may be dramatic.
This scenario (Applegate, Hogan, \& Scherrer 1987)
looked very exciting about ten years ago when it
was noted that the QCD (quark-hadron) transition seemed likely to produce
strong inhomogeneity at just the right distance scale, and
early IBBN calculations indicated a large reduction in $Y_p$
and increase in D/H allowing very large $\eta$, even a critical
density in baryons only.  More detailed calculations showed
that the effects were less dramatic, and the upper limit to $\eta$
given by D/H and $Y_p$ is raised at most by a factor of 2 or 3
as compared to SBBN, and this only if the inhomogeneity was
at near the optimal distance scale ($10^{-3}\ldots10^{-2}$ pc),
and most of the baryon number was in the high density regions.
The most severe problem for this kind of IBBN is ${}^7$Li
overproduction.  Some ${}^7$Li depletion (by a factor of 2 or 3)
in Pop II stars is needed to allow for larger $\eta$ than in SBBN.
Figure 1 is from a recent review of this scenario
by Kainulainen, Kurki-Suonio, \& Sihvola (1999).

Recent lattice QCD calculations favor a much smaller
distance scale, although uncertainties are big enough
so that the optimal distance scale cannot be ruled out.
The distance scale from the electroweak (EW) phase transition
must be so small that the effects on BBN cannot be large; 
in the best case they could be comparable to other small 
effects that have recently been included in accurate BBN codes.

\begin{figure}
\plottwo{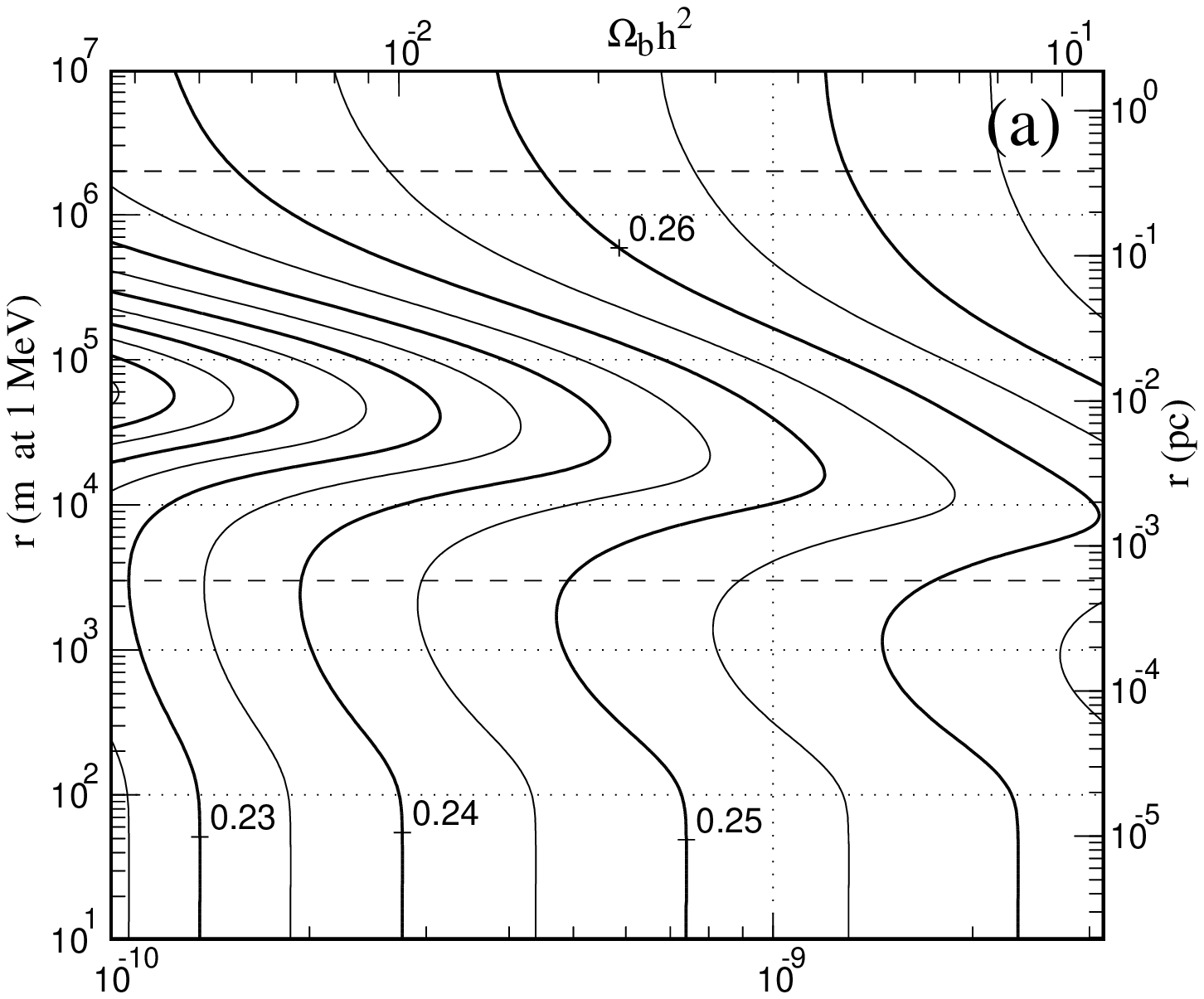}{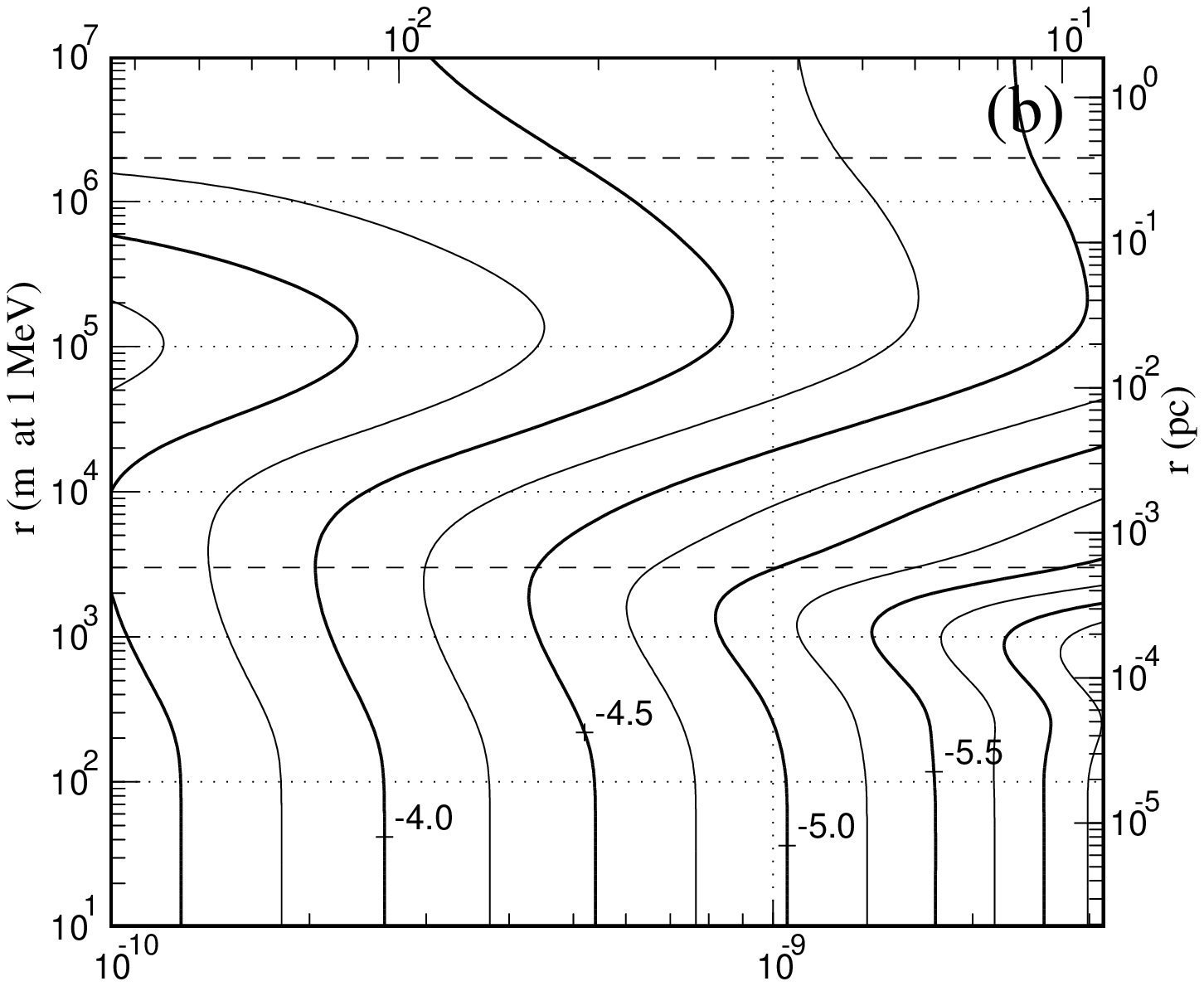}
\plotfiddle{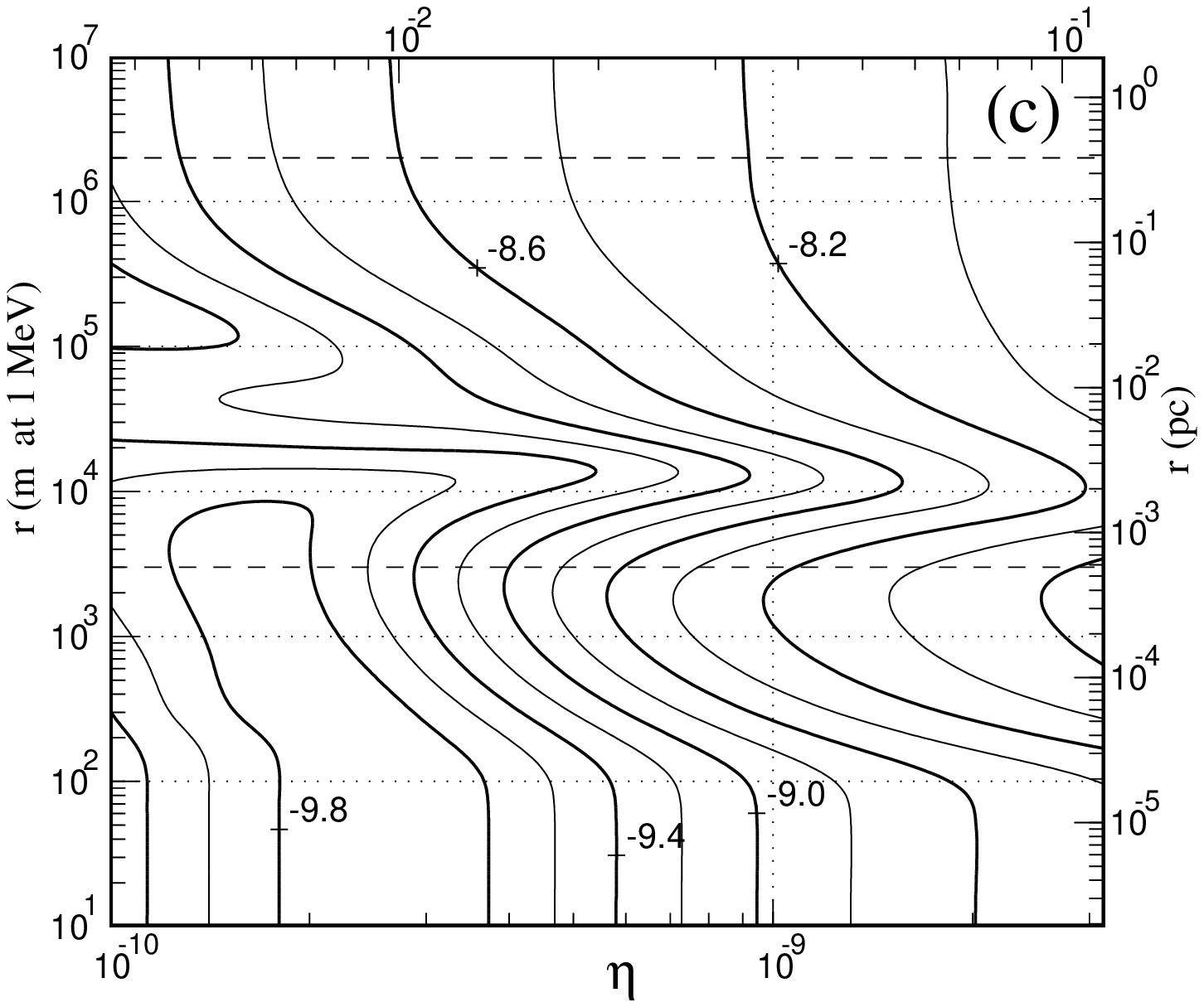}{6cm}{0}{40}{40}{-32}{-66}
\caption{The ${}^4$He, D, and ${}^7$Li yields
from small-scale inhomogeneous nucleosynthesis
runs with a centrally condensed geometry,
with density contrast $R = 800$ and high-density
volume fraction $f_v = 0.125$.
The contours of (a) $Y_p$, (b) $\log_{10}$D/H,
and (c) $\log_{10}{}^7$Li are plotted as
a function of the average baryon-to-photon ratio $\eta$ and the distance scale
$r$ of the inhomogeneity.  The two horizontal dashed lines denote the horizon
scale $\ell_H$ at the QCD (upper) and EW (lower) phase transitions.
From Kainulainen et al. (1999).}
\end{figure}

\subsection{Regions of Antimatter}

A less-studied variant of IBBN is one where $\eta$ is allowed to have
negative values, i.e., there are antimatter regions.  
This is possible in some baryogenesis scenarios (Dolgov 1996).
Antimatter in cosmology has been reviewed by Steigman (1976).
If the distance scale of antimatter regions is small, antimatter
and matter will mix and annihilate in the early universe, and the
presence of matter today implies that there was initially more 
matter than antimatter.  If the distance scale is large, so that
antimatter regions will survive till present, observational constraints
require either the amount of antimatter to be very small, or 
the distance scale to be very large, comparable to the present
horizon or larger (Cohen, De R\'{u}jula, \& Glashow 1998),
so that the case of large regions is not of interest for BBN. 

The smaller the antimatter regions are, the earlier they annihilate.
Rehm \& Jedamzik (1998) considered annihilation immediately before
nucleosynthesis.  Kurki-Suonio \& Sihvola (1999) extended these results
to larger distance scales where annihilation occurs during or after
nucleosynthesis (see Figure 2).  So far the focus has been on
obtaining upper limits to the amount of antimatter at various scales in
the early universe, but clearly there is also potential for obtaining
acceptable abundances with nonstandard values of $\eta$, although
probably only with fine-tuned model parameters.

\begin{figure}
\plottwo{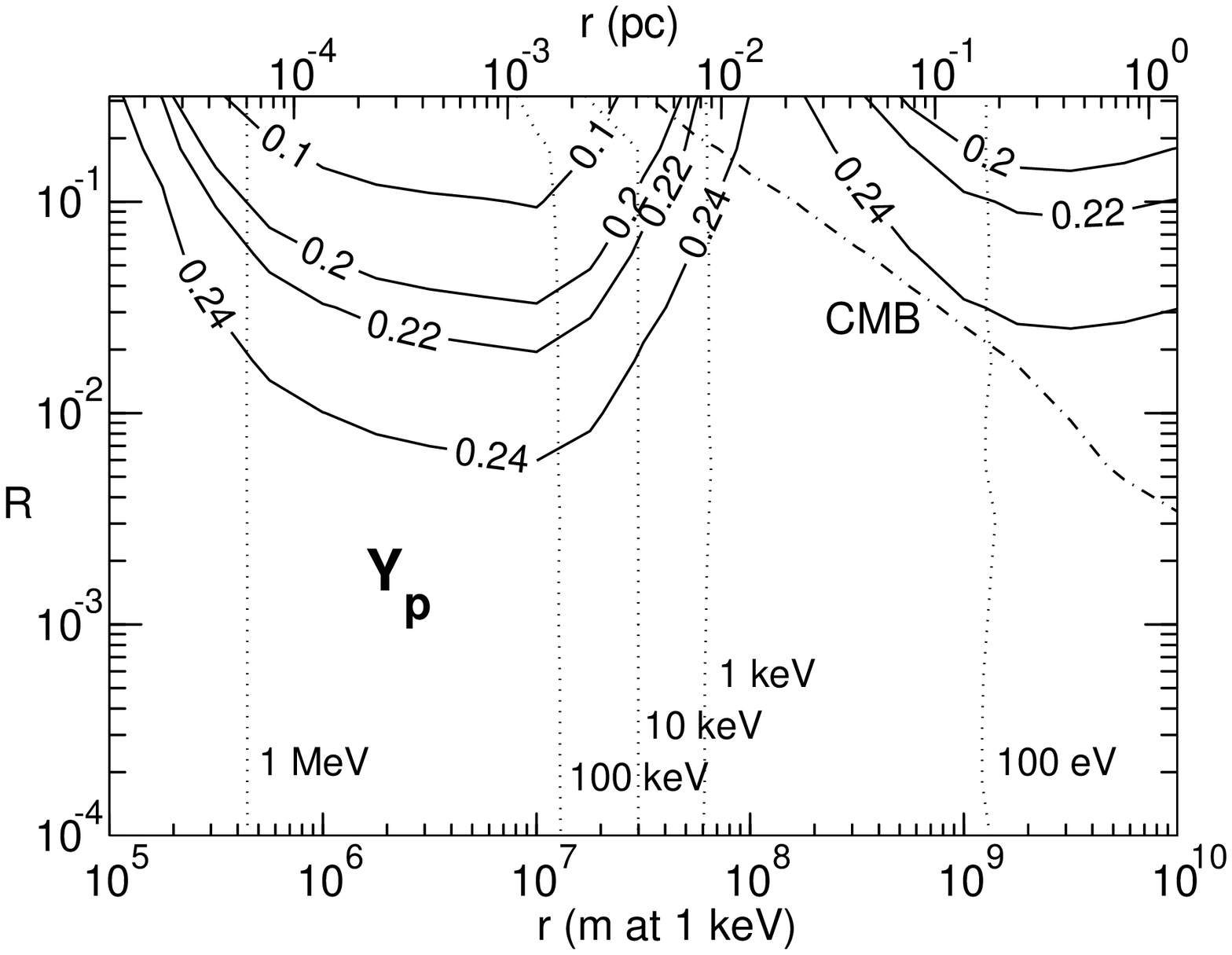}{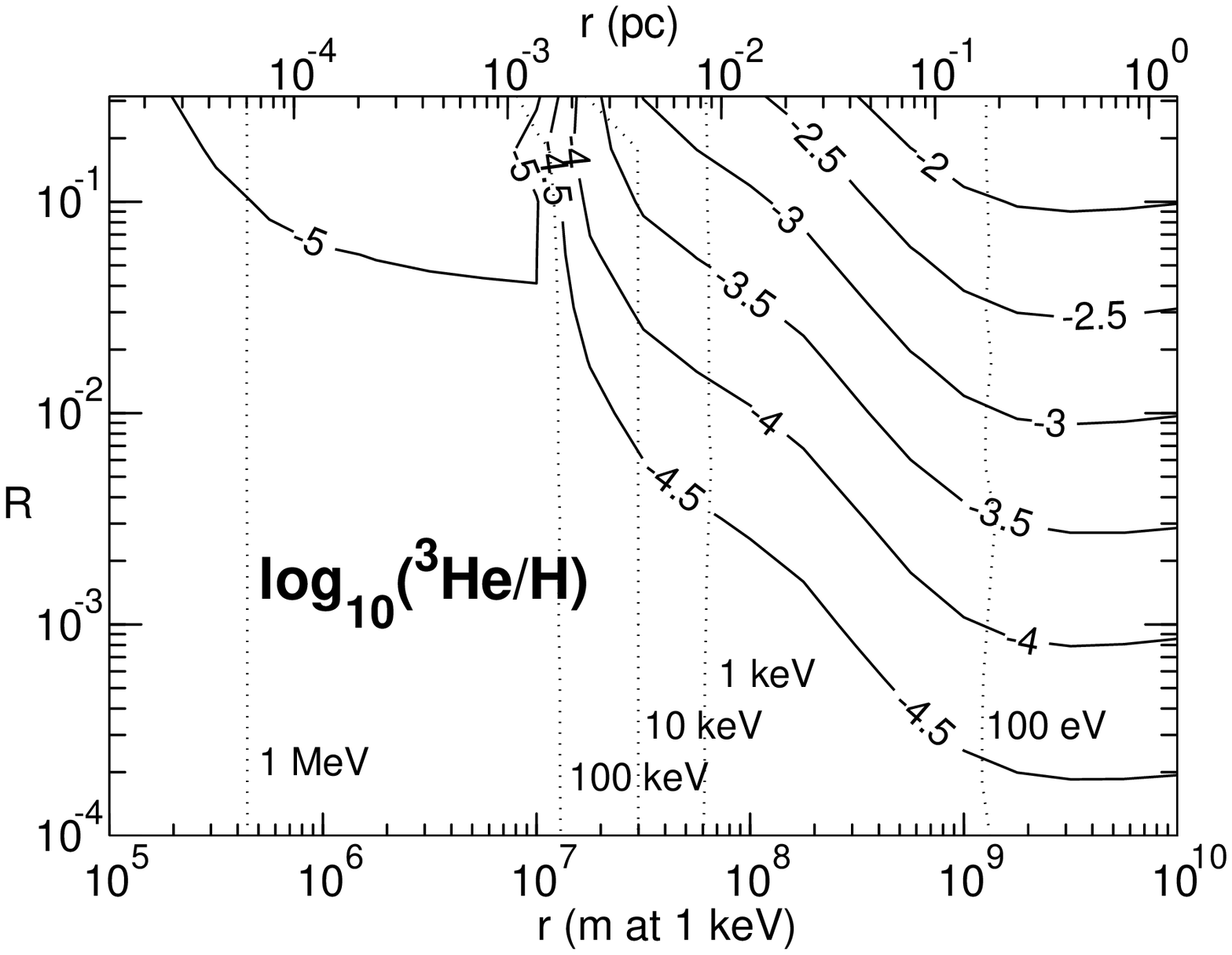}
\caption{The (a) ${}^4$He and (b) ${}^3$He yields as a function of
the antimatter/matter ratio $R$
and the antimatter domain radius $r$.  The distance
scales are given both at $T = 1$ keV (in meters)
and today (in parsecs).
We plot contours of $Y_p$ and (the logarithm of) the
number ratio ${}^3$He/H.  The dotted lines show contours of the
``median annihilation temperature'', i.e., the temperature of the
universe when 50\% of the antimatter has annihilated.  Typically the
annihilation is complete at a temperature lower
than this by about a factor of 3.
The dot-dashed line gives the upper limit to $R$ from CMB spectrum
distortion.  This plot is for $\eta_{10} = 6$.
From Kurki-Suonio \& Sihvola (1999).}
\end{figure}

\section{Neutrinos and Big Bang Nucleosynthesis}

Neutrinos affect BBN in two ways, through the energy density effect
and the $\nu_e$ effect.  The most significant effect is on $Y_p$ in
both cases.

The energy density in neutrinos affects the expansion rate of the
universe.  The simplest way to increase the energy density
of the early universe from the standard model is to have
additional particle species (sterile neutrinos or other hypothetical
particles).  The custom is to parametrize this by an ``effective
number of neutrino species''.  The standard case is $N_\nu = 3$.
We now know that there are only three ``active'' neutrino 
species, so any additional species must be ``sterile'' neutrinos or
other very weakly interacting particles.
A higher energy density means faster expansion.
This leads to $n/p$ freezeout at a higher temperature, leaving more
neutrons, and resulting in a higher
${}^4$He yield.  The D yield is also increased, so an increased
energy density is disfavored by BBN, and one gets an upper limit,
e.g., $N_\nu < 3.2$ (Burles et al. 1999) or
$N_\nu \la 4$ (Lisi, Sarkar, \& Villante 1999),
depending on what observational
constraints one uses.

Electron neutrinos affect the weak $n \leftrightarrow p$ reactions
directly.  More $\nu_e$ leads to fewer neutrons and thus to less
${}^4$He (and everything else), whereas more $\bar{\nu}_e$ leads
to more neutrons and more ${}^4$He.

\subsection{Neutrino Degeneracy}

In SBBN one assumes that the neutrino asymmetry (difference 
between the number of neutrinos and antineutrinos), 
\begin{equation}
   L_\nu \equiv  \frac{n_\nu - n_{\bar{\nu}}}{n_\gamma}
   = 0.069\biggl(\frac{T_\nu}{T}\biggr)^3\bigl(\pi^2\xi + \xi^3),
\end{equation}
which is related to the
neutrino chemical potential $\mu_\nu$, or the degeneracy parameter
$\xi \equiv \mu_\nu/T$, is small, $\ll 1$.  This seems natural,
since the comparable baryon asymmetry $\eta$ is small.  
However, the neutrino background is unobservable, so we cannot
rule out a large neutrino asymmetry.  A  larger asymmetry always 
means a larger neutrino energy density, raising $N_\nu$.
To have a significant effect on BBN, we must have $|\xi|$,
$|L_\nu| \ga 0.1$.
There is a separate contribution from each neutrino flavor.
Thus there are three indepedent
degeneracy parameters, 
$\xi_e$, $\xi_\mu$, and $\xi_\tau$.  The energy density effect
is the same for all three flavors, and depends only on $|\xi|$.
The electron neutrino effect depends only on $\xi_e$, but is much stronger,
and the direction of the effect depends on the sign.  

There are two possible scenarios for affecting BBN.  If
$\xi_e$ is comparable in magnitude to $\xi_\mu$ and $\xi_\tau$,
or larger, one can forget the other
two in first approximation.  One can then adjust $\xi_e$ to dial in the
desired value of $Y_p$.  The other elements are hardly affected.
A less natural scenario is one where the asymmetries in the other
two neutrino flavors are much larger,
and the energy density and $\nu_e$ effects
are balanced against each other to keep $Y_p$ in the acceptable
range.  This way one can have a significant effect on the other 
abundances and raise the acceptable range for $\eta$.  This
second scenario is constrained by structure formation, since the
large neutrino energy density means that the matter/radiation equality
and thus the beginning of structure formation occurs later.
Kang \& Steigman (1992) used a generous lower limit
for matter/radiation equality, $z_{\rm eq} > 10^3$ to widen
the SBBN acceptable range from $\eta_{10}$ = 2.8--4.7 to
$\eta_{10}$ = 2.8--19.

\subsection{Inhomogeneous Neutrino Degeneracy}

The different results from high-$z$ D/H measurements
(Tytler, Fan, \& Burles 1996; Webb et al. 1997) raised the
question whether there might be a large-scale inhomogeneity
in primordial abundances.  This is very difficult to achieve, since
the extreme isotropy of the CMB rules out any significant large-scale
inhomogeneity in $\eta$ or the energy density.
Dolgov \& Pagel (1999) have come up with a way of getting around 
this constraint.  In their model
the asymmetries of the different neutrino flavors are inhomogeneous
but balanced with each other so that they add up to a homogeneous 
total energy density.  The inhomogeneous $\xi_e$ is then responsible
for the inhomogeneous primordial abundances through the $\nu_e$ 
effect.  They suggest that an Afflect-Dine type scenario of
generation of leptonic charge asymmetry, respecting the symmetry
between different lepton families, could be responsible for
creating a domain structure, where the neutrino asymmetries would
have the same three values but interchanged with respect to $e$, $\mu$
and $\tau$.
To achieve a significant D/H inhomogeneity,
a huge $Y_p$ inhomogeneity has to be allowed.
But since there are no high-$z$ $Y_p$ determinations,
this cannot be used to rule out their model.  Table 2 shows
an example of
what kind of abundances we could have in such a domain structure.
The first line would correspond to our local domain; from the
other domains we would have only D/H observations.

\begin{table}
\centering
\caption{Abundances of light elements for $\eta_{10} = 5$ and nonzero
values of all three chemical potentials.  One example from
Dolgov \& Pagel (1999).}
\vspace{2mm}
\begin{tabular}{rrrccc}
\tableline
$\xi_e$ & $\xi_\mu$ & $\xi_\tau$ & D/H & $Y_p$ & ${}^7$Li/H  \\
\tableline
0.1 & $-1$ & 1 & $3.8\times10^{-5}$ & 0.23 & $2.5\times10^{-10}$ \\
$-1$ & 0.1 & 1 & $9.2\times10^{-5}$ & 0.55 & $4.5\times10^{-10}$ \\
1 & $-1$ & 0.1 & $2.8\times10^{-5}$ & 0.08 & $1.1\times10^{-10}$ \\
\tableline
\tableline
\end{tabular}
\end{table}

\subsection{Decay of a Massive Tau Neutrino}

If the rest mass of a neutrino species is much larger than 100 MeV,
then it is becoming nonrelativistic before nucleosynthesis and its
contribution to the energy density is different from the
standard zero-mass case.  The laboratory limits for the neutrino 
masses leave this as a possibility for $\nu_\tau$.  Above
the neutrino decoupling temperature,  $T \sim 3$ MeV, a massive neutrino
species contributes less energy density, because of
neutrino-antineutrino annihilation, but after neutrino
decoupling the annihilation ceases and the rest mass then contributes
extra energy density.
Neutrinos this heavy must decay to avoid contributing
too much to the present energy density.
The decay time and mode
are of crucial importance to BBN.  If the decay time is
very short, then the contribution to $N_\nu$ will be less than one.
The most interesting
case is the one where $\nu_\tau$ decays into $\nu_e$ (and a
scalar particle), since then the $\nu_e$ effect could cause a
significant reduction in $Y_p$.
 
These calculations are difficult since the decisive effects occur
near the neutrino decoupling temperature, so thermal equilibrium is
not maintained and the neutrino spectra are distorted.
The recent results by Hannestad (1998) and Dolgov et al. (1999) are
in disagreement with each other.  Hannestad gets the maximum reduction
of $Y_p$, from the SBBN result $Y_p = 0.239$ to $Y_p < 0.20$, for
$\nu_\tau$ mass $m_\nu$ = 0.2--0.5 MeV and lifetime 
$\tau < 100$ s.  According to Dolgov et al., the maximum reduction is
less, to $Y_p \sim 0.21$, and occurs for larger masses,
$m_\nu$ = 2--3 MeV, and requires a shorter lifetime
$\tau < 1$ s.

The most natural explanation of the SuperKamiokande (1998) result
on atmospheric neutrinos is $\nu_\mu \rightarrow \nu_\tau$
oscillation.  Then $\nu_\tau$ cannot be heavy and its mass
will not affect BBN significantly.  To allow the above 
scenario, the atmospheric neutrino oscillations would have to
be into a sterile neutrino species, $\nu_\mu \rightarrow \nu_s$,
instead (Kainulainen et al. 1999).

\subsection{Neutrino Oscillations}

Observations of solar neutrinos and atmospheric neutrinos
(SuperKamiokande 1998) as
well as the LSND (1998) accelerator experiment see different amounts
of the different neutrino flavors than predicted by the Standard Model.
This can be explained by neutrino oscillations.  This is a
quantum-mechanical phenomenon where the flavor 
$(\nu_e,\nu_\mu,\nu_\tau)$ content of the neutrino varies periodically.
This requires nonzero neutrino masses and the effect is determined
by the difference in mass-squared, $\Delta m^2$, and the ``mixing
angle''.

All three (solar, atmospheric, and LSND)
``neutrino problems'' cannot be simultaneously explained 
by oscillations among three flavors, but require at least a fourth 
flavor, $\nu_s$, which must be ``sterile'', i.e., much more weakly
interacting than the three known ``active'' flavors, in order
not to violate the limit $N_\nu \sim 3$ from $Z^0$ decay width
(Particle Data Group 1998).  A sterile
neutrino would also be useful for supernova nucleosynthesis 
(Peltoniemi 1996; Caldwell, Fuller, \& Qian 1999).

The LSND results are controversial, so the other viewpoint is
to ignore them until they are confirmed by independent experiments,
in which case the solar and atmospheric neutrino problems can be 
explained just with the three active neutrinos.

Oscillations among (light, non-degenerate, i.e., $\xi = 0$) active
neutrinos do not affect BBN,
since they all have equal abundances.  If the sterile neutrino exists,
it would have thermally decoupled from the other neutrinos very early,
much before BBN, so that its contribution to $N_\nu$ would be $\ll 1$.
Active-sterile neutrino oscillations before BBN would
then lead to production of $\nu_s$, increasing  $N_\nu$ 
(Enqvist, Kainulainen, \& Thomson 1992), which from the BBN point
of view is undesirable.  The situation is more complicated, however.
The oscillation depends on the background temperature, and at a
certain temperature there is a resonance.  This resonance temperature
depends on the neutrino energy, so as the temperature falls, the
resonance sweeps through the neutrino spectrum.  If there is a 
small pre-existing asymmetry (this will be the case, since
thermal fluctuations suffice), the rates of neutrino 
and antineutrino oscillation will be different. 
Resonant active--sterile neutrino oscillations will
then lead to a growth of the neutrino asymmetry by a
large factor
(Barbieri \& Dolgov 1991; Foot \& Volkas 1995; Shi 1996;
Enqvist, Kainulainen, \& Sorri 1999; Di Bari \& Foot 2000).
This may generate a large enough electron neutrino asymmetry
to affect BBN (Bell, Foot, \& Volkas 1998; Kirilova \& Chizhov 1998;
Shi, Fuller, \& Abazajian 1999).  

Depending on the oscillation parameters, the asymmetry may
either just grow or
oscillate between positive and negative values, so that the final
sign of the asymmetry becomes unpredictable.
To calculate the effect
on BBN is complicated, since the resulting
distortion of the $\nu_e$ spectrum
is also important for BBN, and the process happens near the neutrino
decoupling temperature.  There are two schemes to generate
a large $\nu_e$ asymmetry, either directly via $\nu_e \leftrightarrow
\nu_s$ oscillations or indirectly via $\nu_{\mu(\tau)} \leftrightarrow
\nu_s$ and $\nu_{\mu(\tau)} \leftrightarrow \nu_e$  oscillations.

This scenario is under active study and there is much
controversy among the different research groups.
In Fig.~3 we show results obtained by Shi et al.
(1999).  The maximal effect on $Y_p$ seems to be
at the $\pm 0.01$ level.

\begin{figure}
\plottwo{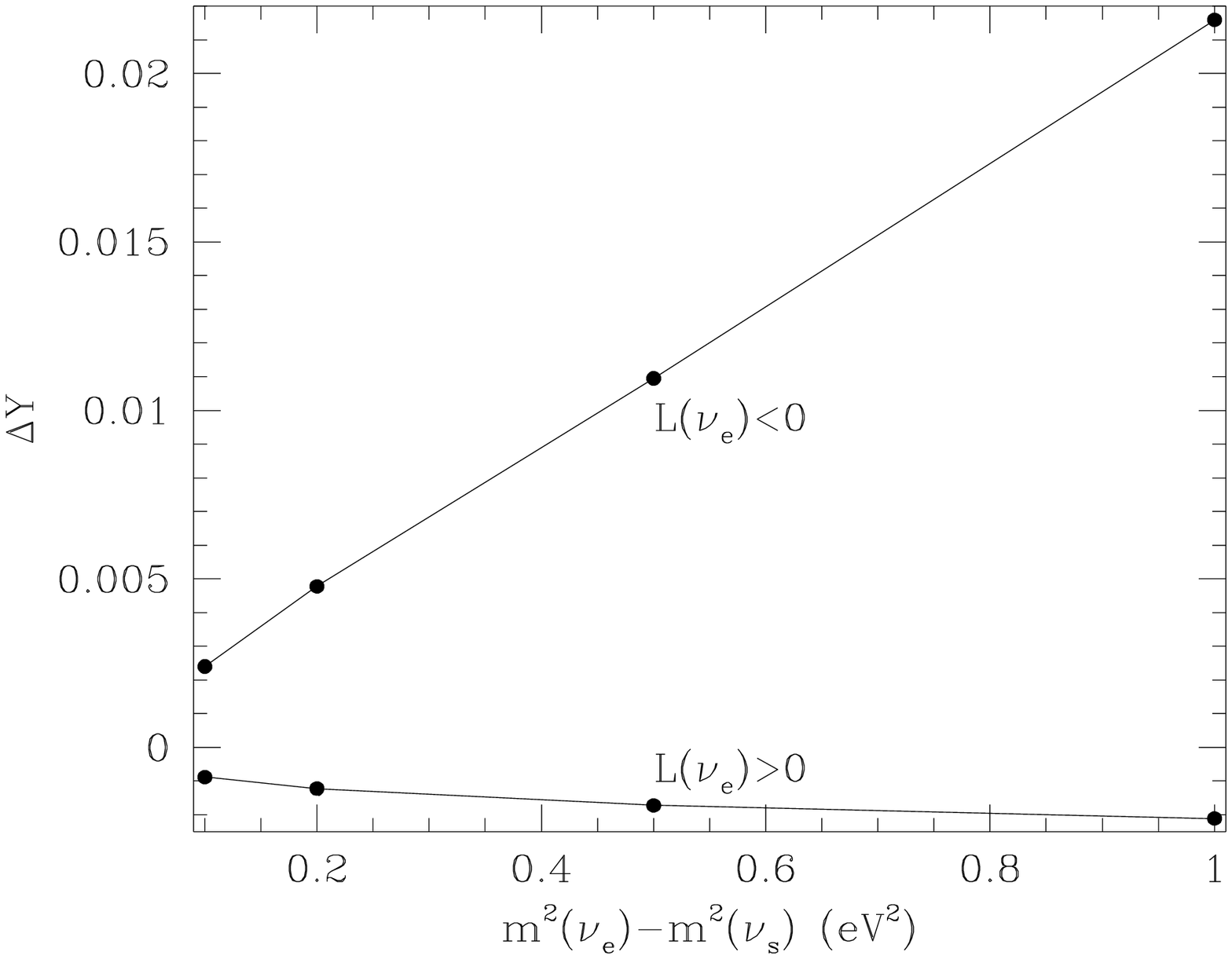}{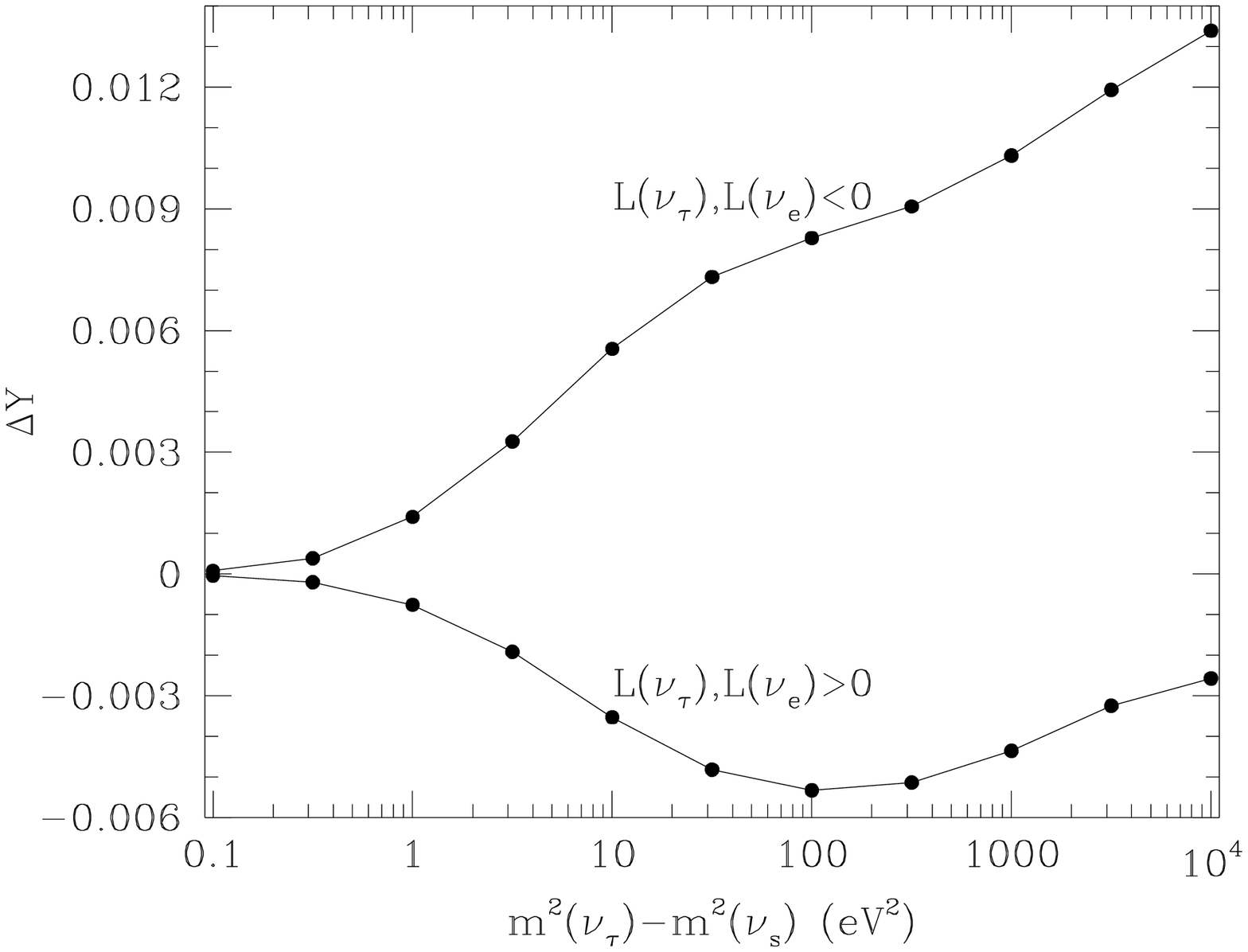}
\caption{The impact on the primordial $^4$He abundance $Y$
if an asymmetry in $\nu_e\bar\nu_e$ is generated by a resonant
$\nu_e\leftrightarrow\nu_s$ mixing (left) or
by the indirect neutrino mixing scheme (right), as a function
of $\Delta m^2$.
The baryon-to-photon ratio is set
to $\eta=5.1\times 10^{-10}$.
From Shi et al. (1999).
}
\end{figure}

\section{Conclusions}

At present, no NSBBN scenario appears as
convincing as SBBN, which is the simplest of all.
Often the real world has turned out to be more complicated in the
end than first assumed, but for the early universe a simple
picture has been very successful.  However, it is healthy to keep
in mind the possibility that SBBN might not be the full story, 
and that any discrepancies between observations and SBBN might
actually be telling us something important about the
early universe or particle physics.

\acknowledgements
I thank K. Kainulainen, A. Kalliom\"{a}ki, J. Peltoniemi, and A. Sorri
for advice on neutrino physics.


\begin{references}
\reference Applegate, J. H., Hogan, C. J., \& Scherrer, R. J. 1987,
 \prd\ 35, 1151
\reference Barbieri, R. \& Dolgov, A. 1991, Nucl.Phys.B 237, 742 
\reference Bell, N. F., Foot, R., \& Volkas, R. R. 1998, \prd\ 58, 105010 
\reference Burbidge, G., \& Hoyle, F. 1998, \apj\ 509, L1
\reference Burles, S., Nollett, K. M., Truran, J. W., \& Turner, M. S.
 1999, \prl\ 82, 4176 
\reference Caldwell, D. O., Fuller, G. M., \& Qian, Y.-Z. 1999,
 astro-ph/9910175
\reference Cohen, A. G., De R\'{u}jula, A., \& Glashow, S. L. 1998,
 \apj\ 495, 539
\reference Di Bari, P. \& Foot, R. 2000, \prd, to be published,
 hep-ph/9912215
\reference Dimopoulos, S., Esmailzadeh, R., Hall, L. J., \& 
 Starkman, G. D. 1988, \apj\ 330, 545
\reference Dolgov, A. D. 1996, in Proceedings of the
 International Workshop on Baryon Instability, Oak Ridge, March 1996,
 hep-ph/9605280
\reference Dolgov, A. D., Hansen, S. H., Pastor, S., \& Semikoz, D. V.
 1999, Nucl.Phys.B 548, 385
\reference Dolgov, A. D., \& Pagel, B. E. J. 1999, New Astron 4, 223
\reference Enqvist, K., Kainulainen, K., \& Thomson, M. 1992,
 Nucl.Phys.B 373, 498
\reference Enqvist, K., Kainulainen, K., \& Sorri. A. 1999, 
 Phys.Lett.B 464, 199
\reference Foot, R. \& Volkas, R. R. 1995, \prl\ 75, 4350
\reference Hannestad, S. 1998, \prd\ 57, 2213
\reference Hata, N., Scherrer, R. J., Steigman, G., Thomas, D., 
 Walker, T. P., Bludman, S., \& Langacker, P.  1995, \prl\ 75, 3977
\reference Kainulainen, K., Kurki-Suonio, H., \& Sihvola, E. 1999,
 \prd\ 59, 083505
\reference Kang, H.-S., \& Steigman, G. 1992, Nucl.Phys.B 372, 494
\reference Kirilova, D. P. \& Chizhov, M. V. 1998, \prd\ 58, 073004
\reference Kurki-Suonio, H., \& Sihvola, E. 1999, astro-ph/9912473
\reference Leonard, R. E., \& Scherrer, R. J. 1996, \apj\ 463, 420
\reference Lisi, E., Sarkar, S., \& Villante, F. L 1999,
 \prd\ 59, 123520
\reference LSND Collaboration (Athanassopoulos, C., et al.) 1998,
 \prl\ 81, 1774; \prc\ 58, 2489
\reference Malaney, R. A., \& Mathews, G. J. 1993, Phys.Rep 229, 145
\reference Particle Data Group (Caso, C., et al.) 1998,
 Eur.J.Phys.C 3, 1
\reference Peltoniemi, J. 1996, in Proceedings of the 3rd Tallinn Symposium
 on Neutrino Physics, ed. I. Ots, J. L\~{o}hmus, P. Helde \& L. Palgi,
 hep-ph/9511323
\reference Rehm, J. B., \& Jedamzik, K. 1998, \prl\ 81, 3307
\reference Sarkar, S. 1996, Rep.Prog.Phys 59, 1493
\reference Shi, X. 1996, \prd\ 54, 2753
\reference Shi, X., Fuller, G. M., \& Abazajian, K. 1999, \prd\ 60, 063002
\reference Steigman, G. 1976, \araa\ 14, 339
\reference Super-Kamiokande Collaboration (Fukuda, Y., et al.) 1998,
 \prl\ 81, 1562
\reference Tytler, D., Fan, X. M., \& Burles, S. 1996, 
 Nature 381, 207
\reference Webb, J. K., Carswell, R. F., Lanzetta, K. M., Ferlet, R., 
 Lemoine, M., Vidal-Madjar, A., \& Bowen, D. V. 1997, Nature 388, 250
\end{references}
\end{document}